\documentclass{emulateapj}
\usepackage{natbib}
\begin{document}
\slugcomment{Submitted to The Astrophysical Journal}
\title{Ensemble analysis of open cluster transit surveys: upper limits on the frequency of short-period planets consistent with the field}

\author{Jennifer L. van Saders and B. Scott Gaudi}
\affil{Department of Astronomy, The Ohio State University, 140 West 18th Avenue, Columbus, OH, 43210}
\email{vansaders@astronomy.ohio-state.edu}

\shortauthors{van Saders \& Gaudi}
\shorttitle{Limits on the planet frequency in open clusters}
\begin{abstract}

Several photometric surveys for short-period transiting giant planets have targeted a number of open clusters, but no convincing detections have been made. Although each individual survey typically targeted an insufficient number of stars  to expect a detection assuming the frequency of short-period giant planets found in surveys of field stars, we ask whether the lack of detections from the ensemble of open cluster surveys is inconsistent with expectations from the field planet population. We select a subset of existing transit surveys with well-defined selection criteria and quantified detection efficiencies, and statistically combine their null results to show that the upper limit on the planet fraction is $5.5\%$ and $1.4\%$ for 1.0 $R_{J}$ and 1.5 $R_{J}$ planets, respectively in the $3<P<5$ day period range. For the period range of $1<P<3$ days we find upper limits of $1.4\%$ and $0.31\%$ for 1.0 $R_{J}$ and 1.5 $R_{J}$, respectively. Comparing these results to the frequency of short-period giant planets around field stars in both radial velocity and transit surveys, we conclude that there is no evidence to suggest that open clusters support a fundamentally different planet population than field stars given the available data.

\end{abstract}

\section{Introduction}
The unexpected discovery of giant planets orbiting their host stars with periods of a few days \citep{mayor1995, butler1997} made it clear that our theories of planet formation and evolution are incomplete \citep[e.g.][]{lin1996}. Although progress has been made toward understanding the the presence of short-period giant planets, many details of their formation and evolution remain uncertain. 

That short-period planets form from circumstellar disks is clear, but the exact mechanism in which disk material condenses to form planets has yet to be established. Jupiter mass planets are thought to form through one of two mechanisms: core accretion \citep{mizuno1980, bodenheimer1986, pollack1996}, in which a rocky or icy core reaches a critical mass and rapidly accretes an envelope of gas, or gravitational instability \citep{boss1997}, in which the massive, cool disk fragments and collapses into giant planets. Core accretion theories have historically had difficulty producing planets of the correct masses and locations of the giant planets in our solar system within a $\sim5$ Myr protoplanetary disk lifetime \citep[e.g.][]{pollack1996}, although recent work has been more successful \citep[see][and references therein]{movshovitz2010}. Gravitational instability, on the other hand, requires relatively cool, quiescent disk conditions and sufficiently short cooling timescales, both of which may not be physically realistic at the radii at which gas giants are thought to form \citep{rafikov2005}.

It is clear that short-period giant planets cannot form \emph{in situ}, since protoplanetary disks at a few tenths of an AU are too hot and do not contain sufficient mass to produce a gas giant. Therefore, in addition to the uncertainties regarding the formation mechanism, we must face additional uncertainties related to the means by which a planet migrates from its hypothesized birthsite in order to explain gas giants in short-period orbits. In what is labeled Type I migration \citep{ward1997, goldreich1980}, the planet experiences torques from the massive gaseous disk, a loss of angular momentum, and subsequent inward motion toward the star. Sufficiently massive planets are capable of opening a gap in the circumstellar disk, which can also result in inward drift as the disk accretes onto the star; this is called Type II migration \citep{lin1986, ward1997}. Recently, planet migration through planet-planet scattering and eccentricity pumping due to a stellar binary companion coupled with tidal dissipation (``Kozai migration'') have also been proposed as migration mechanisms \citep[see][]{nagasawa2008,wu2003,mazeh1997}. The relative importance of each of these mechanisms to the process of planetary migration remains unclear.

Each of these different formation mechanisms should imprint certain signatures onto the resulting planet population. One hopes, with a sufficiently large observed sample of planets, to be able to determine the relative importance of each of these formation channels. From an observational point-of-view, the way to understand the origin of short-period planets is to search a large number of stars for planets and produce a correspondingly rich sample of exoplanets. The statistical properties of these planets and their observed frequency can then help to constrain models of planet formation and evolution. It is therefore essential that any sample of stars searched for planets has well characterized properties so that trends among the stars that are found to host planets can be recognized.

\citet{janes1996} suggested that open clusters were ideal targets for planet searches since they host such a uniform and easily characterizable population of stars. In contrast to field stars, the metallicity, age, and distance to cluster stars are relatively easy to determine. As such, open clusters represent a potential testbed for planet formation theories. The frequency of short-period planets provides information about how often gas giants form, migrate, and survive in a single, uniform, coeval population of stars. Furthermore, observations of several clusters of different metallicities allow us to probe the planet-metallicity correlation \citep{fischer2005}, in which higher metallicity field stars are more likely to host a planet.

Planet search surveys in clusters also allow us to characterize the degree to which the cluster environment itself impacts the formation and survival of planetary systems. The cluster properties at birth and throughout its evolution are critical to determining whether the cluster environment is capable of affecting the formation and survival of planetary systems. The degree to which the effects of close encounters and photoevaporating radiation present in dense stellar environments influences the formation, evolution, and survival of planets are not well understood. There is evidence that the observed stellar densities of old open clusters today are significantly less than they were at birth, and that the current environments in old open clusters are not what they were during the first 5-10 Myr when giant planets were presumably forming. In particular, a given cluster may lose 10-80\% of its stars in the process of emerging from its embedded stage, and thus only the initially most massive clusters survive and are observed as old open clusters today \citep[see][]{lada2003, friel1995}.

The degree to which planetary systems are affected by the cluster environment is still poorly understood both observationally and theoretically. Theoretical investigations of the importance of environment \citep{kobayashi2001, adams2006, bonnell2001,armitage2000} suggest that open clusters are not dense enough to significantly affect the formation and survival rates of planets, especially within the $\sim$ 5-30 AU in which gas giants that later migrate are thought to form. However,  observations of the Orion Nebula Cluster (ONC) \citep{eisner2008} found that less than $10\%$ of the stars harbor disks of mass comparable to that of the minimum mass solar nebula, and that the frequency of disks decreased with proximity to the massive central Trapezium stars. They also claim that the frequency of disks in the dense ONC is statistically different from that of the lower density Taurus cluster. A similar study of the Arches also detected a disk fraction that varies with proximity to the massive, central stars \citep{stolte2010}.

Of the numerous planet detection methods available to astronomers, photometric transit searches are among the most logical choices for an open cluster survey \citep{janes1996, pepper2005,pepper2006, vonbraun2005}. Transit surveys have the ability to monitor a large number of stars simultaneously, and benefit from the fact that cluster stars are concentrated over a small portion of the sky. Photometric surveys can also be performed on smaller, more easily accessible telescopes than competing methods of planet detection \citep{pepper2005, pepper2006}. The photometric data are simple to calibrate, and in the case of clusters the metallicities, distance, ages, and radii for all stars can be obtained with relative ease. Because the technological requirements for a successful photometric survey are fulfilled by modest and readily available instruments, transit searches are a simple and direct method of detecting planets in these systems, with the caveat that the radial velocity follow-up needed to confirm planet candidates is not always as easily obtained \citep[see][]{aigrain2007}.

However, despite a number of transit searches targeting open clusters \citep{hartman2009, burke2006, mochejska2008, mochejska2005, mochejska2006, miller2008, bramich2005,bramich2006, hood2005, bruntt2003, rosvick2006, vonbraun2005, hidas2005, montalto2007, howell2005, pepper2008, vonbraun2004, vonbraun2005, lee2004, hidas2005, street2003, rosvick2006, montalto2009}, no convincing planetary transits have been found. There are two possible explanations for this paucity of planets \citep{janes2009}: either there is something different about the planet populations of open clusters, or there are simply too few observed cluster stars to detect a planet from an otherwise ``typical'' planet population. The frequency of short-period planets is $\sim1\%$ \citep{cumming2008}, and the geometric probability that such a planet transits is only $\sim10\%$, meaning that one naively expects to observe $\sim1000$ stars before detecting even one transiting planet. Because the transit depths are small ($\sim1\%$) and have a low duty cycle, both precise (to a few milimagnitudes) photometry and excellent temporal coverage are required in any transit search. When one considers the need for sufficient a signal-to-noise ratio and multiple observed transits for robust detection, it is generally the case that many more than 1000 stars must be monitored to detect any transiting planets \citep{burke2006, vonbraun2005}. Since the richest open clusters contain a few thousand stars at best, the lack of detected transiting planets in any individual survey may simply be due to an insufficient number of target stars.

While each survey alone may not observe a sufficient number of stars to have expected a planet detection, we estimate the total number of cluster stars observed in all transit surveys to be roughly 10,000, and that combined the sample of stars is rich enough to derive interesting upper limits on the frequency of short-period planets. By carefully statistically combining the null results of several photometric surveys, we provide a more stringent upper limit than is possible with individual studies, and then compare this limit to the frequencies of planets among field stars derived from both photometric and radial velocity surveys. As we will show, the typical detection efficiencies of these photometric surveys are insufficient, even with 10,000 total stars, to place tighter constraints on the frequency of short-period planets than surveys of field stars, and thus we conclude that these combined upper limits are consistent with the observed frequency of short-period giant planets in the field. Given the available data, we have no reason to suspect that the population of short-period planets in open clusters is anything other than ``ordinary''.

The paper is organized as follows: in \S\ref{sec:methods} we discuss the manner in which upper limits on planet frequencies are derived from transit surveys and how their individual results can be combined, in \S\ref{sec:surveys} we discuss our selection of the transit surveys included in our analysis and the normalization of individual survey results. In \S\ref{sec:results} we present our combined upper limits. \S\ref{sec:discussion} and \S\ref{sec:conclusion} are devoted to discussion and our conclusions, respectively.

\section{Methods}\label{sec:methods}
\subsection{Finding planets in a photometric survey}
The techniques used to find planets in any photometric survey are broadly similar. All surveys attempt to achieve high photometric accuracy with maximal temporal coverage over the longest possible period of time. The better the temporal coverage, the more sensitive a survey is to all transits, and the longer the duration of the survey, the larger the range of planetary periods that are detectable. We refer the reader to \citet{pepper2005} and \citet{vonbraun2005} for extensive discussions of the factors that determine a survey's sensitivity to transits and the design of successful surveys.

Once the data have been collected, one produces light curves and searches for periodic variability. The box-fitting least-squares (BLS) method introduced by \citet{kovacs2002} is a popular means to do so within the transit community. The algorithm searches for periodic, rectangular deviations from a flat light curve, and is an objective, repeatable means of identifying transit candidates. Conversely, some authors may choose to identify transits by eye. Regardless of the method used to identify photometric eclipses, the planetary nature of convincing transit candidates must then be confirmed with radial velocity follow-up observations, since astrophysical false positives can mimic the transit of a Jupiter-sized object.

In the case of a null result, some authors choose to carefully quantify their detection efficiencies and place upper limits on the frequency of short-period planets. One typically injects a large number of simulated, limb-darkened transits into constant simulated stars or light curves from the survey itself, and then subjects these transits to the same detection algorithms with the same selection criteria as the original data. The fraction of injected transits that are recovered quantifies the detection efficiency. This efficiency is a complicated combination of effects that is discussed in more detail in the next section.

While the general process is similar for all surveys, the details of how each author chooses to perform the steps varies slightly, and in practice it is not necessarily correct to directly compare or combine the results of two surveys. Authors may make different assumptions about the period distribution of planets, or quote their results for differing planetary radii, etc., all of which are important when comparing the results of multiple different surveys. In order to combine the results of several surveys, we must re-normalize, to the extent that we can, to a common set of assumptions.

\subsection{Quantifying upper limits}
In this section we discuss the mathematical description of the detection efficiency of a transit survey, the calculation of upper limits on the frequency of planets using a null result, and the method for combining several normalized surveys into a single upper limit.

In general, the expected number of planets detected in the radius range $R_p$ to $R_p +dR_p$ and period range $P$ to $P+dP$ can be written \citep{burke2006}

\begin{equation}
\displaystyle \frac{d^2N_{exp}}{dR_pdP} = f_p \frac{d^2P}{dR_pd\mathcal{P}} \sum_k^{N_{\star}}{\mathcal{P}_{mem,k}\mathcal{P}_{tr,k}\mathcal{P}_{det,k}},
\label{eqn:diff}
\end{equation}
where $k$ is the star, the sum is over all $N_{\star}$ stars, and $\mathcal{P}_{mem,k}$ is the probability that the star is a cluster member. $f_p$ is the fraction of stars that host short-period planets distributed as $d^2\mathcal{P}/dR_pdP$. We assume $d^2\mathcal{P}/dR_pdP$ is independent of the stellar mass. This distribution is poorly constrained by exoplanet statistics, and is usually assumed as a prior. We will assume that this distribution is a delta function in the radius and uniform in the logarithm of the period:

\begin{equation}
\displaystyle \frac{d^2\mathcal{P}}{dR_pdP} \propto \displaystyle \frac {\delta(R_p - R_p^{'})}{P},
\end{equation}
where $R_p^{'}$ is the planetary radius of interest. $\mathcal{P}_{tr,k}$ is the geometric probability that a planet transits its host star, such that
\begin{equation}
\mathcal{P}_{tr,k} = \displaystyle\frac{R_p+R_{\star}}{a} \propto M_{\star}^{-1/3}(R_p+R_{\star})P^{-2/3},
\end{equation}
where $R_{\star}$ is the radius of the star and $a$ is the semimajor axis of the orbit, where we have assumed the orbit is circular.$\mathcal{P}_{det,k}$ is the probability that a planet of radius $R_p$ and period $P$ will be detected around star $k$ if it transits, given the precise temporal coverage, precision, and signal-to-noise ratio of the observations. In practice $\mathcal{P}_{tr}$ is trivial to calculate, while $\mathcal{P}_{det}$ is emphatically not so. It is, in general, the single most difficult factor to characterize in the entire formalism because it represents a complex interplay between numerous different observational effects. Simple analytic estimates of $\mathcal{P}_{det}$ tend to overestimate the probability that the survey will be capable of detecting transits \citep{beatty2008}. The best way to characterize $\mathcal{P}_{det}$ is to perform Monte Carlo simulations \emph{a posteriori} with transits injected into the light curves of constant stars observed in the survey or simulated constant stars.

When authors choose to perform an analysis of the detection efficiencies with Monte Carlo simulations, the resulting efficiencies generally account for $\mathcal{P}_{det}$,  $\mathcal{P}_{tr}$, and $d^2\mathcal{P}/dR_pdP$. By injecting transits with randomly sampled parameters from an assumed period distribution into stars in the survey that were searched for transits, one automatically accounts for the intrinsic mass function of stars and planet period distributions. Factors such as noise present in the light curves and the observing window, which are crucial to determining $\mathcal{P}_{det}$, are also automatically included in transit injection and recovery schemes.  $\mathcal{P}_{mem}$ is generally determined separately through proper motion measurements, proximity of a star to the cluster main sequence, off-cluster comparison fields, or modeling of the field star population to determine contamination levels. After the detection efficiencies are quantified, one has

\begin{equation}
\displaystyle N_{exp} = f_p \int_{R_{p_{min}}}^{R_{p_{max}}} \int_{P_{min}}^{P_{max}}\frac{d^2\mathcal{P}}{dR_pdP}\sum_{k}^{N_{\star}}\mathcal{P}_{mem,k}\mathcal{P}_{tr,k}\mathcal{P}_{det,k}
\label{eqn:Ndet}
\end{equation}
where $N_{exp}$ is the number of transiting planets that are expected to be detected, and $f_p$ is the only unknown factor. By assuming that $\mathcal{P}_{det,k}$ is independent from $\mathcal{P}_{mem,k}$ for each star we can define an average membership probability $\langle\mathcal{P}_{mem}\rangle$. While these two factors are not strictly uncorrelated, it is a reasonable approximation and simplifies the analysis considerably. Using this assumption we can average over all of the stars in a survey such that
\begin{equation}
 N_{exp}=N_{\star}f_p\langle\mathcal{P}_{mem}\rangle\langle\mathcal{P}_{\epsilon}\rangle,
\label{eqn:pepsilon}
\end{equation}
where $\langle\mathcal{P}_{\epsilon}\rangle$ is the average probability over all stars that a planet will both transit \emph{and} be detected.

In the case in which no planets are detected over the course of the survey, one can derive an upper limit on the frequency of planets of a given radius within a specific period range. Formally, the probability of seeing $N_{det}$ successful detections in $N_{\star}$ independent trials, each with a probability of success $f_p\langle{\mathcal{P}_{\epsilon}}\rangle\langle{\mathcal{P}_{mem}}\rangle$ is given by the binomial distribution
\begin{eqnarray}
&&\displaystyle \mathcal{P}\left(N_{det};N_{\star},f_p\langle\mathcal{P}_{\epsilon}\rangle \langle\mathcal{P}_{mem}\rangle\right) = \\
&&{N_{\star} \choose N_{det}}
\left({f_p\langle\mathcal{P}_{\epsilon}\rangle\langle{\mathcal{P}_{mem}}\rangle}\right)^{N_{det}}\left(1-f_p\langle\mathcal{P}_{\epsilon}\rangle\langle\mathcal{P}_{mem}\rangle\right)^{N_{\star}-N_{det}} \nonumber
\end{eqnarray}
Because the number of stars (and therefore trials) is large and the probability of detecting a planet around a given star is small, we can assume the Poisson approximation to the binomial distribution without compromising our results. The probability of detecting $N_{det}$ events when the number of expected events is $N_{exp}$ is
\begin{equation}
\mathcal{P}(N_{det};N_{exp})=\displaystyle \frac{N_{exp}^{N_{det}}e^{-N_{exp}}}{N_{det}!}.
\end{equation}
In the case of a non-detection, $N_{det}=0$ and
\begin{equation}
\mathcal{P}(0;N_{exp})=e^{-N_{exp}}.
\end{equation}
As mentioned earlier, the number of detections we expect for a given survey goes as
\begin{equation}
N_{exp}=f_p N_{\star}\langle\mathcal{P}_{mem}\rangle\langle\mathcal{P}_{\epsilon}\rangle = f_{p}N_{cl}\langle\mathcal{P}_{\epsilon}\rangle,
\end{equation}
where $N_{cl} \equiv N_{\star}\langle\mathcal{P}_{mem}\rangle$ is the number of cluster members observed. The 95\% confidence upper limit on the planet fraction (when $\mathcal{P}(0;N_{exp})=0.05$) is then
\begin{equation}
f_p \leq \displaystyle \frac{-\ln{\left(0.05\right)}}{N_{cl}\langle\mathcal{P}_{\epsilon}\rangle} \approx \frac{3}{N_{cl}\langle\mathcal{P}_{\epsilon}\rangle}.
\label{eqn:fp}
\end{equation}

The multiple surveys that we utilize in this paper each quote a value of $f_p$ or $N_{cl}\langle\mathcal{P}_{\epsilon}\rangle$ specific to their observations, chosen planetary radii, and period ranges. Provided the limits or detection efficiencies for all surveys are quoted for the same planetary radii and period ranges, the upper limits from individual surveys can be combined using
\begin{equation}
 f_{tot} \leq \frac{-\ln{(0.05)}}{\sum_s{N_{cl,s}\langle\mathcal{P}_{\epsilon,s}\rangle}}\textrm{,}
\end{equation}
where $f_{tot}$ is the combined upper limit on the planet fraction, and $s$ indexes individual surveys. The expression for $f_{tot}$ can be written simply as
\begin{equation}
f_{tot} \leq \displaystyle \left(\sum_s{\frac{1}{f_s}}\right)^{-1}.
\label{eqn:ftot}
\end{equation}

\section{Selection of Transit Surveys and Basic Assumptions}\label{sec:surveys}

Although the general planet search techniques and subsequent quantification of detection efficiencies are relatively similar among the photometric surveys considered here, there are always slightly different assumptions made during the process by different authors. In the transit surveys we ultimately selected for our study, the two main areas in which differences were apparent were the assumptions made about the period distribution of planets and the estimates of cluster membership. Such differences are relatively straightforward to re-normalize, and do not pose a fundamental obstacle to combining the surveys. However, in the cases in which authors do not quantify their detection efficiency sufficiently, or do so only in terms of the most simplistic analytic estimates, we cannot re-normalize their results, and therefore cannot include them here. We also have no way to quantify the detection efficiency of transit surveys in which the transits were identified ``by eye'', and so these too cannot be utilized.

We choose transit surveys with the following characteristics that make it possible to re-normalize and combine their results:

\begin{enumerate}
\item{Transits are detected through the use of an automatic transit detection algorithm with rigorous, defined thresholds of detection. Specifically,``by eye'' transit detection is not sufficient, since detection probabilities are not quantifiable.}
\item{A simulation of detection efficiency was conducted using constant stars, and injected transits are limb-darkened and have a representative sampling of periods, inclinations, and planetary radii. This allows for a realistic quantification of elements such as the window of observations, instrument noise, and the effectiveness of the transit detection algorithms.  }
\item{Survey results must be accompanied by an estimate of cluster membership probabilities. It is essential that we have a reasonable estimate of the actual number of observed cluster stars, as opposed to a blind mixture of cluster and field stars. Without a membership estimate, any upper limits derived from the survey results for the frequency of planets in clusters is not meaningful. Surveys are included in cases in which membership is not explicitly estimated but sufficient information to do so is available.}
\end{enumerate}

Our analysis therefore includes the results from 6 separate transit surveys of 6 open clusters: M37 \citep{hartman2009}, NGC 1245 \citep{burke2006}, NGC 188 \citep[hereafter M08]{mochejska2008}, NGC 6791 \citep[hereafter M05]{mochejska2005}, NGC 2158 \citep[hereafter M06]{mochejska2006}, and NGC 2362 \citep{miller2008}. In the case of NGC 6791, three separate surveys targeted the cluster and we choose to utilize only one for simplicity. The fundamental cluster parameters (distance, age, etc.) of the systems observed in the relevant surveys are given in Table \ref{cluster_params} and a discussion of the individual surveys follows in \S\ref{sec:individual_clusters}.

\begin{deluxetable*}{lllllllll}
\tabletypesize{\footnotesize}
\tablecolumns{7}
\tablewidth{0pt}
\tablecaption{Fundamental cluster parameters}
\tablehead{
                        \colhead{Cluster}&
                        \multicolumn{2}{c}{$(\alpha^{\circ}, \delta^{\circ})$}&
                        \multicolumn{2}{c}{$(l,b)$}&
                        \colhead{distance (pc)}&
                        \colhead{age (Myr)}&
                        \colhead{[Fe/H]} &
                        \colhead{References\tablenotemark{a}}}
\startdata
M37      & $88.074$  & $+32.553$   & $177.637,$  & $+3.094$   & $1500\pm100$  & $550\pm30$        & \phn$0.05\pm0.04$  & 2,2,2\\
NGC 188  & $12.108$  & $+85.255$   & $122.865,$  & $+22.384$  & $1710\pm80$   & ${6000^{+1000}}$  & $-0.04\pm0.05$     & 7,8,6 \\
NGC 1245 & $48.68$   & $ +47.25$   & $146.64,$   & $-8.92$    & $2800\pm200$  & $1040\pm90$       & $-0.05\pm0.09$     & 1,1,1 \\
NGC 2158 & $91.054$  & $+24.097$   & $186.634,$  & $+1.781$   & $3600\pm400$  & $2000\pm300$      & $-0.06$            & 5,5,10 \\
NGC 2362 & $109.65$  & $ -24.98$   & $238.20,$   & $-5.58$    & $1480$        & $5^{+1}_{-2}$     & \nodata            & 3,3,- \\
NGC 6791 & $290.221$ & $ +37.772$  & $69.958,$   & $+10.904$  & $4200$        & $8000\pm500$      & $0.39\pm0.01$      & 4,4,9 \\
\enddata
\tablenotetext{a}{References for cluster parameters, given in the order: distance, age, metallicity. References are as follows: 1.) \citet{burke2004}, 2.) \citet{hartman2008}, 3.) \citet{moitinho2001}, 4.) \citet{chaboyer1999}, 5.) \citet{carraro2002}, 6.) \citet{vonhippel1998}, 7.) \citet{sarajedini1999}, 8.) \citet{dinescu1995}, 9.) \citet{carraro2006}, 10.) \citet{jacobson2009}         }

\label{cluster_params}
\end{deluxetable*}

In addition, we adopt the following conventions to normalize the results from the six surveys:
\begin{enumerate}
\item{We assume that the underlying planet population is an even logarithmic distribution in period. In cases where the authors of the papers have not made a similar assumption, we correct for the difference between the assumed distributions.}
\item{We quote upper limits at the 95\% confidence level, unless otherwise stated.}
\item{ ``Hot Jupiters'' (HJs) are defined to be planets with periods in the range $3<P<5$ days, and ``Very Hot Jupiters'' (VHJs) those that have periods of $1<P<3$ days.}
\end{enumerate}

\subsection{Planet Candidates}

Some of the selected surveys detected transit-like events of the appropriate depth among their light-curves, which represent candidate transiting planets. We assume here that all candidates are false positives. We justify this assumption based on the fact that most candidates are revealed as false positives upon closer inspection \citep[see][]{brown2003,evans2010}, none of these candidates have been confirmed as planets through radial velocity follow-up observations, and, as we will describe later in \S\ref{sec:discussion}, our upper limits are consistent with radial velocity and transit surveys of field stars even without the inclusion of the candidate planets (and thus our assumptions are conservative).

\section{Individual Cluster Surveys}\label{sec:individual_clusters}
The upper individual limits derived for each cluster and the details of the re-normalization of each survey follow. Our final adopted upper limits derived from individual surveys are listed in Table \ref{upper_limits}.

\begin{deluxetable*}{lllllllll}
\tabletypesize{\footnotesize}
\tablecolumns{9}
\tablewidth{0pt}
\tablecaption{Upper limits on the frequency of HJs and VHJs}
\tablehead{
                       \colhead{}&
                       \colhead{}&
                       \colhead{}&
                       \multicolumn{3}{c}{VHJ}&
                       \multicolumn{3}{c}{HJ}\\
                       \colhead{Cluster}&
                       \colhead{$N_{cl}$}&
                       \colhead{}&
                       \colhead{1.0 $R_J$}&
                       \colhead{1.2 $R_J$}&
                       \colhead{1.5 $R_J$}&
                       \colhead{1.0 $R_J$}&
                       \colhead{1.2 $R_J$}&
                       \colhead{1.5 $R_J$ }    }
\startdata

M37 & 1450 & & 0.027 & 0.025 & 0.023 & 0.083 &0.077 & 0.069\\
NGC 188 & 797 & firm & 0.222 & 0.160 & 0.067 & 1.87 & 1.34 & 0.560\\
 &   & marginal & 0.051 & 0.042 & 0.027 & 0.178 & 0.144 & 0.093\\
NGC 1245 & 870 & & 0.240 & 0.170 & 0.064 & -- & -- & 0.360\\
NGC 2158 & 2460 & firm & 0.076 & 0.049 & 0.009 & 0.394 & 0.255 & 0.047\\
 &   & marginal &0.021 & 0.015 & 0.006 & 0.052 & 0.038 & 0.016\\
NGC 2362 & 475 & & 0.521 & 0.369 & 0.142 & 4.99 & 3.30 & 0.775\\
NGC 6791 & 1997 & firm & 0.082 & 0.053 & 0.008 & 0.346 & 0.221 & 0.033\\
 & & marginal & 0.025 & 0.018 & 0.006 & 0.069 & 0.049 & 0.018\\

\enddata
\tablenotetext{.}{Please note that $N_{cl}$ refers to the estimated number of cluster members observed with sufficiently precise photometry to detect planets. }
\label{upper_limits}
\end{deluxetable*}

\subsubsection{M37}

The \citet{hartman2009} survey of M37 utilized injected, limb-darkened transits with randomly selected periods, radii, and inclination angles to arrive at their detection probabilities, and determined cluster membership by using the luminosity function of a nearby blank field and a narrow strip on the CMD enclosing the cluster main sequence to arrive at  membership probabilities as a function of $r$ magnitude.  Planets are assumed to be evenly distributed in $\log{P}$ and $\cos{i}$. We adopt the authors' quoted upper limits without adjustment.

\subsubsection{NGC 188, NGC 2158, and NGC 6791}

The M08 survey of NGC 188, M06 survey of NGC 2158, and M05 survey of NGC 6791 all utilize identical transit injection and recovery schemes to ascertain their detection efficiencies. A total of 432,000 limb-darkened transits are injected into observed light curves without transits for each survey and run through the detection algorithms for recovery. They inject signals corresponding to periods from 1.05 to 9.85 days in linear steps of 0.2 days, radii from 0.95 to 1.50 $R_{J}$ in steps of 0.05 $R_{J}$, and $\cos{i}$ from 0.0125 to 0.9875 in steps of 0.025. Detections are defined as ``marginal'' or ``firm'' by the authors if they fulfill one or both of the transit selection criteria: the period of the injected transit was recovered to within 2\% of the true value or that of an alias, and certain thresholds were met for the values of the BLS statistics. We consider both classes of detections here, but use the ``firm'' detections in all combined upper limits quoted later in the text and in all plots, since a ``marginal'' detection can represent an event with only a single detected transit.

There are two ways in which we must adjust the published values in order to compare them to other surveys in our sample. The authors assume that planets are distributed linearly in period, while we have adopted the convention that the distribution is uniform in the logarithm of the period, and so we must rescale their results to reflect our choice. Secondly, estimates of cluster membership are lacking or in need of fine tuning, and we therefore estimate membership probabilities based on the information provided in each of the three papers.

In all three papers the authors assume that planets are uniformly distributed in period. The average probability of detecting a planet of any of the considered radii (0.95 to 1.50 $R_{J}$) and inclinations ($\cos{i}$ from 0.0125 to 0.9875) is given by the period-frequency histograms in Fig. 4 in M08, Fig. 6 in M06, and Fig. 5 in M05. In order to rescale the detection efficiencies to a planet distribution uniform in the logarithm of the period, we apply a multiplicative scaling factor to each period-frequency histogram bin. This scaling factor, $S_{P_{dist}}$, is given by

\begin{equation}
S_{P_{dist},j}=\displaystyle \frac{P_{max}-P_{min}}{P_{bin,j}\ln{\left(\frac{P_{max}}{P_{min}}\right)}},
\end{equation}
where $j$ is the bin index, $P_{max}\left(P_{min}\right)$ is the largest (smallest) period in the period range of interest (i.e. 1 and 3 days, corresponding to HJs with $1<P<3$ days), and $P_{bin,j}$ the period upon which the histogram bin is centered. The net effect of this correction is to increase the detection probability of short-period planets relative to that of longer period planets.

Each period-frequency histogram bin is an \emph{average} over all planetary radii and inclinations. If we wish to calculate the probability of detection for a particular radius, we must multiply the bins by yet another scaling factor that quantifies whether a transit of a particular planetary radius is detected more or less efficiently than average.  For this adjustment we use the radius-frequency histograms in Fig. 5 M08, Fig. 8 in M06, and Fig. 7 M05, which plot the average probability of detecting a planet of a particular radius, where the probabilities represent an average over all considered periods and inclinations. The scaling factor is given by

\begin{equation}
 \displaystyle S_{r}=\frac{\mathcal{P}_{R_p}}{\langle{\mathcal{P}\rangle}},
\end{equation}
where $\mathcal{P}_{R_p}$ is the probability, read from the radius-frequency histogram, of detecting a planet of radius $R_p$, and $\langle{\mathcal{P}\rangle}$ is the average detection probability over all radii.

With these scaling factors we can then calculate the detection efficiency for any desired period range and planet radius, $\mathcal{P}_{\epsilon}$ from Eqn. \ref{eqn:pepsilon}, described as

\begin{equation}
 \langle\mathcal{P}_{\epsilon}\rangle=\displaystyle \frac{\displaystyle S_{r}\sum_{j=1}^{N_{bins}}{\mathcal{P}_{j}S_{P_{dist},j}\Delta{P}}}{P_{max}-P_{min}},
\end{equation}
where $\mathcal{P}_{j}$ is the probability of detection in each period bin in the frequency-period histogram, $N_{bins}$ is the number of bins contained in the period range $P_{min}<P<P_{max}$, and $\Delta{P}$ is the period width of each individual bin. This technique is applied to all three of the clusters NGC 188, NGC 2158 and NGC 6791.

\subsubsection{NGC 6791 membership}
As noted earlier, it is critical that we have estimates of the actual number of cluster stars observed. The M05 paper provides estimates the number of cluster and field stars in the sample used for the transit analysis. Candidate cluster members are selected as those stars that lie within 0.06 mag of the $V-R$ vs.$R$ cluster main sequence. While this significantly reduces the field star contamination, there is still a non-negligible number of field stars among these candidate cluster members that happen to have the same colors as the cluster main sequence. To quantify this contamination we use the Besanc\c{o}n stellar population synthesis galaxy model \citep{robin2003} to estimate the number of field stars that should be present in the region of the cluster CMD used to determine membership. Secondly, to confirm the result of the first method, we then estimate the typical field star density on the CMD and check that the resulting membership is consistent with the estimates derived from the galaxy model.

 We first utilized the Besanc\c{o}n galaxy model to estimate the number of field stars expected in the M05 field of NGC 6791. The models were run for a field of view of 0.144 deg$^{2}$ with 25 dust clouds evenly spaced every 2pc from the observer, each with an extinction of 0.019 for a total $A_{V}=0.477$ from \citet{schlegel1998}. We found that this yielded more realistic star counts than using a uniform diffuse extinction over the entire distance to the cluster. We restrict the sample in the same manner as M05 to stars with $R$ magnitudes $17<R<19.8$, which corresponds to all stars below the main sequence turnoff and an rms uncertainty in the $R$ magnitude in the M05 survey of less than 0.05 mag. All stars from the Besanc\c{o}n simulation within 0.06 mag of the cluster main sequence were then selected as field star contaminants. 1523 of the 5225 field stars predicted by the model were within this region. The M05 member selection method therefore would have found 1523 cluster and 3702 fields stars in this model data.  It is important to note that we do not trust the absolute star counts produced by the model, but expect the fractional number of stars that have colors similar to the cluster main sequence to be more accurate. The ratio of stars detected off of the main sequence in M05 and the simulation is $2871/3702 \approx 0.78$, and thus M05 sees $\sim22\%$ fewer stars that are more than 0.06 mag from the MS than the simulations of the same field. We scale the number of stars present in the model near the MS to the ratio of the number of stars detected in the field portions of both the simulated and actual cluster CMD. We would therefore expect that $1523 \times 0.78 = 1181$ stars selected as cluster members in M05 are actually field star contaminants. This implies that only 1997 of 3178 candidate cluster stars are likely to be true members, so $N_{cl} = 1997$.

Another simple, and somewhat crude, method of estimating the field star contamination is to simply find the average density of stars on the cluster CMD near, but not on, the main sequence. This density, multiplied by the area covered by the main sequence (in mag$^2$), yields an estimate of the number of field stars contaminating the selected sample of cluster stars. We find surface densities of 650-700 stars mag$^{-2}$ and thus the main sequence area of $0.12 \times 3.54$ mag$^2$ yields roughly 300 contaminating stars along the MS strip on the single-field CMD in Figure 2 of M05. This would correspond to roughly 1200 stars over all four observed fields, and about 2000 true cluster members. While the method is approximate at best, it yields a reassuringly similar result. Finally, \citet{kalunzy1992} find that the cluster CMD is subject to 30$\%$ contamination by field stars for stars with $V<20$ mag in NGC 6791, which is again comparable to our estimates.

\subsubsection{NCG 188 membership}
M08 does not provide any estimate of the number of cluster stars observed in NGC 188. However, \citet[hereafter P03]{platais2003} provides a catalog of stars down to $V=21$ in NGC 188, and obtained proper motions and membership probabilities using measurements from earlier epochs taken with photographic plates. These data cover a magnitude range in $V$ similar to that of the stars utilized for the transit search in M08. Since the P03 field only covers a fraction of the field of the M08 photometric search, we use the central field (labeled A3 and B1 in M08) of the M08 survey to derive estimates of the field star density across all 6 of the other observed fields. This reference field is centered on the cluster and $\sim450$ stars are detected in M08. This same field in the P03 survey contains 612 stars, all but 6 of which have measured individual cluster membership probabilities. Of these $\sim600$ stars, 372 stars have membership probabilities $P_{mem}>50\%$, and 234 stars have $P_{mem}<50\%$. To be conservative, we assume that all stars with $P_{mem}<50\%$ are field stars. For a reference field size of $11'.4\times11'.4$ this implies a field star density of $\sim6500$ field stars deg$^{-2}$ down to $V\sim21$. However, since M08 detected fewer stars in this same field, we scale the P03 field star density by $\frac{461}{612}$, the ratio of stars detected  in the M08 and P03 surveys respectively. This gives a field star density of $\frac{461}{612} \times 6482 \textrm{ stars/deg}^{-2} = 4883$ field stars deg$^{-2}$. M08 covers 7 $11'.4\times11'.4$ fields, so we expect 1234 field stars. Since 2031 stars are detected on all 7 chips combined, this suggests that 1234 are field stars, and the remaining 797 are cluster members, and so $N_{cl} = 797$. We note that the earlier estimates that made use of the Besanc\c{o}n model are not feasible here, since we have no estimate of the number of field stars with which to calibrate the star counts returned by the model.

\subsubsection{NGC 2158 membership}

The M06 analysis of NGC 2158 also did not include any estimate of the number of cluster members observed. In this case, existing surveys of the cluster contained either no membership information, covered too small a field of view to be easily comparable, or did not cover the relevant magnitude ranges. To be conservative, then, we make the assumption that M06 field 1 centered on ${\alpha \approx 91.65^{\circ}},{ \delta \approx 23.9^{\circ}}$ contains only field stars. This is a $11^{'}.4 \times 11^{'}.4$ region centered roughly 16' from the cluster center. \citet{kharchenko1997} estimates the maximum radius of the cluster to be $\sim15'$, so this field likely contains at least some cluster stars, although the steep radial density profile of the cluster \citep[see Fig. 6 in][]{kharchenko1997} suggests it will be minimal. Note also that field 1 in M06 is also the most distant field of the four from nearby M35, which produces a slight gradient in the star counts across the M06 fields. M06 finds 675 stars in this field with photometry of suitable quality to be used in the search for transits. If we assume all of these stars are field stars it implies that there are a total of 2700 field stars across all four fields. A total 5159 stars, both cluster and field, with suitable photometry were found in all four fields. We can then estimate $N_{cl}$ as $5159-2700 \approx 2460 $. This is a conservative estimate of the fraction of the observed stars that are true cluster members, since it is unlikely that all 675 stars were truly field stars. This approach preserves the fidelity of the upper limits we derive using this value.

\subsubsection{NGC 1245}
The \citet{burke2006} survey of NGC 1245 includes detailed detection probability simulations and careful treatment of cluster membership. Membership probabilities are determined using a star's proximity on the color magnitude diagram of NGC 1245 to the best fit isochrone. Because each individual star is assigned a separate membership probability, we do not correct further for field star contamination as we did in the M05 NGC 6791 survey. Planets are assumed to be uniform in the logarithm of the semimajor axis, which is equivalent, up to an irrelevant normalization constant, to a distribution uniform in $\log{P}$. As such, the upper limits on the planet fraction as derived by \citet{burke2006} are used in our analysis without adjustment, save for upper limit on 1.5 $R_{J}$ companions in the HJ period range, which we discuss below.

\citet{burke2006} takes HJs to the be planets with periods $3.0<P<9.0$ days, while we have assumed the the HJ period range to be $3.0<P<5.0$ days. The authors quote an upper limit $f_p < 36\%$ for the period range $3.0<P<6.0$ days for 1.5 $R_J$ planets. One can see from the middle panel in Fig. 8 of their paper that the upper limits for a period range of $3.0<P<5.0$ days will be negligibly different than the quoted values for $3.0<P<6.0$ days. If anything, the constraints on the slightly longer period range should be weaker than those of the $3<P<5$ day range, and so we are conservative in adopting the value of $f_p < 0.36$ for the $3<P<5$ day 1.5 $R_J$ planets. No limit could be placed on the frequency of 1.0 $R_J$ planets in this period range.

\subsubsection{NGC 2362}

As with the other surveys, the \citet{miller2008} analysis of NGC 2362 light curves relies on objective and clearly defined thresholds of detection and a transit injection scheme to characterize the sensitivity of the survey. A rough estimate of cluster membership from \citet{irwin2008} using both field star counts from the Besanc\c{o}n galaxy model and simple radial distribution arguments arrives at a field star contamination of roughly $60\%$ over the magnitude range of stars utilized in the transit search.

The authors quote upper limits on planet frequency for the $1<P<3$ days and $3<P<10$ days period ranges, while here we are interested in the $3<P<5$ range. We use their Figure 7 to estimate the fraction of recoveries in the 3-10 day range that would also be detected in the 3-5 day range. Using both the 2 transit and 3 transit curves we estimate that $f_{3-5}\approx0.58\times f_{3-10}$. Combined with the number of expected detections in their Table 7, this corresponds to $f_p \le 4.99$ for 1.0 $R_J$ planets and $f_p \le 0.76$ for 1.5 $R_J$ planets at 95\% confidence in the 3-5 days period range.  Note that $f_p\ge 1.0$ simply means that the data is only able to place limits on the fraction of stars with more than one planet in a given period range.

\section{Upper limits on the planet frequency}\label{sec:results}

With the upper limits on the planet frequencies for each of the transit surveys in hand we can provide a combined upper limit using Eqn. \ref{eqn:ftot} in \S\ref{sec:methods}. Upper limits are presented for planetary radii of 1.0 and 1.5 $R_{J}$ in the HJ and VHJ period ranges here and in Table \ref{fractions}. We find upper limits to the fraction of planets with radii of 1.0 $R_{J}$ and 1.5 $R_{J}$ to be 5.5\% and 1.4\%, respectively. We find upper limits on the frequency of VHJ of 1.4\% and 0.31\% for 1.0 $R_{J}$ and 1.5 $R_{J}$ respectively.

In general, the average radius of short-period planets is greater than 1.0$R_J$ and less than 1.5$R_J$, with mean and median in the range $\sim1.2-1.3$ $R_J$ (from www.exoplanets.org, as of June 2010), and thus to compare to field surveys for planets, we would prefer to quote upper limits for this radius range explicitly. However, very few authors explicitly calculate upper limits for these radii, and thus we must interpolate. If we assume that the detection probability increases linearly with increasing planet radius (as Figures 7, 8, and 5 in M05, M06, and M08, respectively imply), we can also estimate upper limits for 1.2 $R_J$ objects. For this linear interpolation, the upper limits are 0.95\% and 3.9\% for the VHJ and HJ period ranges, respectively.

Unfortunately, the vast majority of surveys fail to provide errorbars on their upper limits or expected number of detections. In addition, the nature of our membership probabilities are such that they represent estimates only. For both of these reasons we do not quote uncertainties on our upper limits.

\begin{figure}
      \centerline{\includegraphics[scale=0.5]{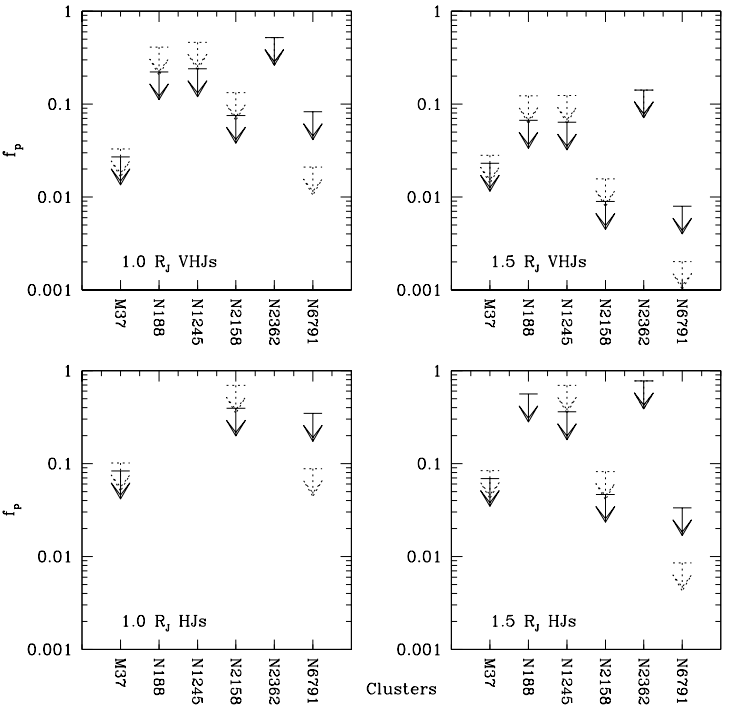}}
       \caption{Upper limits at 95\% confidence for each individual cluster, for both HJ and VHJ period ranges and radii 1.0 $R_J$ and 1.5 $R_J$. Solid arrows represent upper limits derived without any rescaling for the cluster metallicity, and the dashed arrows are rescaled to $\langle\textrm{[Fe/H]}\rangle = 0.09$ using the FV05 relation between host star metallicity and planet frequency. In cases for which the upper limit is formally $f_p \ge 1.0$, we have plotted the upper limits as $f_p = 1.0$. }
\label{fig:indiv_survey_ul}
\end{figure}

\subsection{Metallicity considerations}

There is a well known correlation between the metallicity of the host star and the likelihood that the star bears a planet that can further inform our analysis. \citet{gonzalez1997} first noted that planets tend to be observed around stars with higher than average metallicity. If this correlation is due to the fact that  primordial clouds of higher metallicity more efficiently form planets, rather than pollution due to the process of planet formation itself \citep{santos2000,santos2001,laughlin2000}, we would expect to see relatively more planets in open clusters of higher metallicity, and give more weight to the null results of transit surveys of more metal-rich clusters. \citet{fischer2005} provides the empirical relationship \citep[for a recently updated version see][]{johnson2010} between the probability of hosting a planet and the metallicity of the host:
\begin{equation}
\mathcal{P}( \textrm{[Fe/H]})=0.03 \times 10^{2.0\textrm{[Fe/H]}}.
\end{equation}
The relationship is derived from a sample of 850 stars with uniform planet detectability in Keck, Lick and Anglo-Australian Observatory radial velocity surveys, and is valid for planet periods of less than 4 yrs, velocity amplitudes $K>30$ m/s, and $-0.5< \textrm{[Fe/H]}<0.5$. The weighted average metallicity of all of the cluster stars (not including those from NGC 2362) used in our analysis is $\textrm{[Fe/H]} = 0.09$. The cluster NGC 2362 does not appear to have any metallicity estimate present in the literature, and so we have assigned it the average metallicity of the remainder of our sample.

Cluster metallicities are subject to large uncertainties in the literature, and different authors may publish metallicities discrepant by up to $\sim0.5$ dex in [Fe/H]. In the case of NGC 2158, literature estimates range from $-0.64\pm0.24$ \citep{janes1979} to $-0.238\pm0.064$ \citep{twarog1997} to $-0.03\pm0.14$ \citep{jacobson2009}. We adopt the \citet{jacobson2009} value. Likewise, estimates of the metallicity of NGC 6791 typically range in the literature from $\textrm{[Fe/H]=+0.35-0.45}$ \citep{gratton2006,origlia2006,anthony-twarog2007,carraro2006}. We adopt a metallicity for NGC 6791 of $\textrm{[Fe/H]}=+0.39$ from \citet{carraro2006}. In addition to the significant discrepancies among literature values, cluster metallicities are often quoted without uncertainties. We therefore stress that while it is useful to include our knowledge of the planet-metallicity correlation our analysis, the resulting upper limits should be considered within the context of the uncertainties.

\begin{deluxetable*}{lllllllll}
\tabletypesize{\footnotesize}
\tablecolumns{9}
\tablewidth{0pt}
\tablecaption{Upper limits on the frequency of HJs and VHJs normalized to $\textrm{[Fe/H]}= 0.09$}
\tablehead{
                       \colhead{}&
                       \colhead{}&
                       \colhead{}&
                       \multicolumn{3}{c}{VHJ}&
                       \multicolumn{3}{c}{HJ}\\
                       \colhead{Cluster}&
                       \colhead{$S_{\textrm{[Fe/H]}}$}&
                       \colhead{}&
                       \colhead{1.0 $R_J$}&
                       \colhead{1.2 $R_J$}&
                       \colhead{1.5 $R_J$}&
                       \colhead{1.0 $R_J$}&
                       \colhead{1.2 $R_J$}&
                       \colhead{1.5 $R_J$\tablenotemark{.}}            }
\startdata

M37 &      0.82& & 0.033 & 0.031 & 0.028 & 0.101 &0.094 & 0.084\\
NGC 188 &      0.54 & firm & 0.409 & 0.295 & 0.123 & 3.44 & 2.48 & 1.03\\
 & & marginal & 0.095 & 0.077 & 0.049 & 0.328 & 0.265 & 0.171\\
NGC1245 &      0.52& & 0.463 & 0.327 & 0.123 & -- & -- & 0.695\\
NGC 2158 &      0.57& firm & 0.133 & 0.086 & 0.016 & 0.693 & 0.449 & 0.082\\
 & & marginal &0.036 & 0.026 & 0.011 & 0.092 & 0.066 & 0.028\\
NGC2362 &       1.00& & 0.521 & 0.369 & 0.142 & 4.99 & 3.30 & 0.775\\
NGC 6791 &       3.93& firm & 0.021 & 0.013 & 0.002 & 0.088 & 0.056 & 0.008\\
 & & marginal & 0.006 & 0.004 & 0.002 & 0.018 & 0.012 & 0.005\\

\enddata
\label{upper_limits_met}
\end{deluxetable*}

We derive scaling factors that account for the planet metallicity correlation:
\begin{equation}
 S_{\textrm{[Fe/H]}}=\frac{\mathcal{P}\left(\textrm{[Fe/H]}\right)}{\mathcal{P}\left(0.09\right)},
\end{equation}
which alters the form of Eqn. \ref{eqn:fp} to
\begin{equation}
 f_p \leq \displaystyle \frac{-\ln{\left(0.05\right)}}{\displaystyle S_{\textrm{[Fe/H]}}N_{cl}\mathcal{P}_{\epsilon}},
\end{equation}
since the probability of detecting a planet around a given star is enhanced or depreciated by a factor of $S_{\textrm{[Fe/H]}}$ in comparison to the average star in the sample. Such a scaling means that the null results of metal-rich clusters like NGC 6791 become more important in the calculation of the combined upper limit.

Values of $S_{\textrm{[Fe/H]}}$ for each cluster are as follows: 0.82 for M37, 0.54 for NGC 188, 0.52 for NGC 1245, 0.57 for NGC 2158, 1.0 for NGC 2362, and 3.93 for NGC 6791. In the case of NGC 2362, we have assumed that it has the average metallicity of the entire sample, and $f_p$ is therefore unchanged by the inclusion of metallicity information. The upper limits to the planet fractions adjusted for metallicity are given in Table \ref{upper_limits_met}. When combined, these values yield an upper limit on the planet frequency of 4.3\% for 1.0 $R_{J}$ HJs, 2.9\% for 1.2 $R_{J}$ HJs, 0.69\% for 1.5 $R_{J}$ HJs, 1.1\% for 1.0 $R_{J}$ VHJs, 0.72\% for 1.2 $R_{J}$ VHJs, and 0.16\% for 1.5 $R_{J}$ VHJs, with all values quoted here normalized to $\textrm{[Fe/H]}=0.09$.

\begin{deluxetable*}{llllllll}
\tabletypesize{\footnotesize}
\tablecolumns{8}
\tablewidth{0pt}
\tablecaption{Combined upper limits on $f_p$}
\tablehead{
                       \colhead{$R_J$}&
                       \colhead{Period}&
                       \multicolumn{2}{c}{w/o [Fe/H] correction}&
                       \multicolumn{2}{c}{$\langle\textrm{[Fe/H]}\rangle = 0.09$\tablenotemark{a}} &
                       \colhead{$\langle\textrm{[Fe/H]}\rangle = 0.006$} &
                       \colhead{$\langle\textrm{[Fe/H]}\rangle = -0.084$} \\
                       \colhead{}&
                       \colhead{}&
                       \colhead{firm}&
                       \colhead{marginal}&
                       \colhead{firm}&
                       \colhead{marginal} &
                       \colhead{} &
                       \colhead{$90\%$ c.l.}                   }
\startdata
1.0 & VHJ & 0.0137 & 0.0066 & 0.0109 & 0.0044 &0.0073 &0.0037\\
1.2 & VHJ & 0.0095 & 0.0049 & 0.0072 & 0.0031 & 0.0048 & 0.0032\\
1.5 & VHJ & 0.0031 & 0.0024 & 0.0016 & 0.0013 & 0.0011 & 0.0006\\
1.0 & HJ & 0.0549 & 0.0194 & 0.0431 & 0.0123&0.0290&0.0148\\
1.2 & HJ & 0.0385 & 0.0143 & 0.0286 & 0.0088 & 0.0193 & 0.0127\\
1.5 & HJ & 0.0139 & 0.0068 & 0.0069 & 0.0036 & 0.0046 & 0.0024\\

\enddata
\tablenotetext{a}{$\textrm{[Fe/H]} = 0.09$ corresponds to the average metallicity of the cluster stars, $\textrm{[Fe/H]} = 0.006$ to that of stars in RV planet searches, and $\textrm{[Fe/H]} = -0.084$ to the stars observed in OGLE.}
\label{fractions}
\end{deluxetable*}

\section{Discussion}\label{sec:discussion}

\subsection{Comparison to other planet search results}
\begin{figure*}
	\centerline{\includegraphics[scale = 0.75, trim = -25 300 0 0]{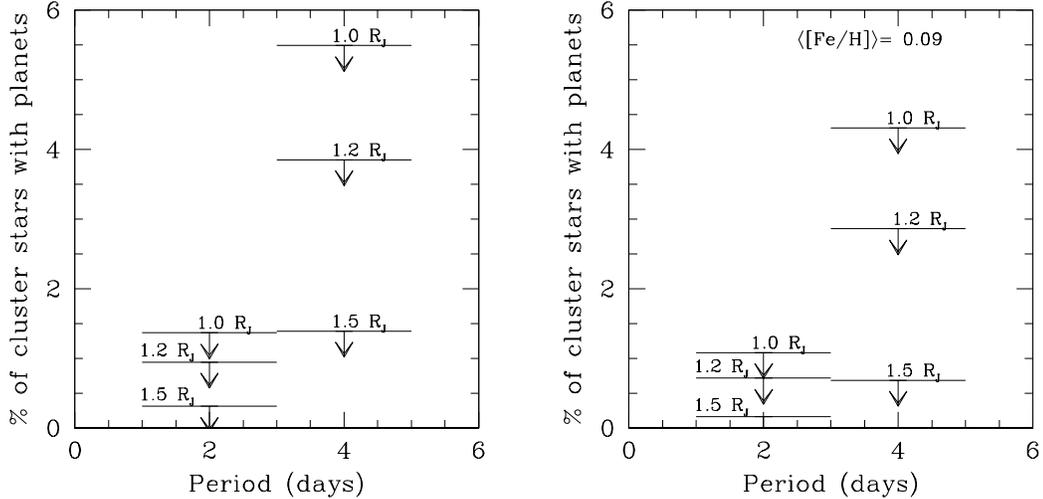}}
       \caption{Combined upper limits utilizing the six selected surveys. The left panel shows upper limits derived without metallicity rescaling for 1.0, 1.2, and 1.5 $R_J$ planets. The upper limits in the right right panel are rescaled based on the cluster metallicities according to the FV05 planet-metallicity correlation to the average metallicity of stars in the sample $\langle\textrm{[Fe/H]}\rangle = 0.09$.}
\end{figure*}

The frequency of short-period planets has been previously estimated in a number of other surveys.  \citet[hereafter G06]{gould2006} used the OGLE III microlensing surveys and found the frequency of short-period planets in the Galactic field (at 90$\%$ confidence) to be $\sim0.31\%$ for $3<P<5$ days and  $0.14\%$ for $1<P<3$ days, valid for planetary radii of $1.0<R_{J}<1.25$. The \citet[hereafter C08]{cumming2008} analysis of the radial velocity Keck Planet Search results suggest that 7/585 of FKG stars have companions with mass $M\sin{i} \ge 0.1$ $M_J$ and periods $3<P<5$ (see Fig. 5 in C08) and 1/585 stars have planets in the period range of $1<P<3$ days. \citet{weldrake2008} found upper limits on the frequency at 95\% confidence of $0.67\%$ for $3<P<5$ days and $0.096\%$ for $1<P<3$ days in the globular cluster $\omega$ Centauri. The \citet{paulson2004} radial velocity survey of the Hyades open cluster detected no close in giant planets among 94 stars, although upper limits on the planet frequency were not calculated.

In comparing the results of other planet searches to the upper limits we calculate for open clusters we must be aware of biases inherent in each survey method. Different survey methods are particularly sensitive to different planetary parameters, and may not always probe the same underlying population. One must also be careful to account for the effects of metallicity of the population of stars surveyed in each study. This is particularly true in the case of radial velocity surveys, which tend to be biased towards more metal rich stars, and in which planet detectability is determined by the mass as opposed to the radius of the planet.

The latter effect is primarily an issue when one wishes to compare radial velocity survey results to transit surveys. The detectability of a planet in a transit survey is not a function of its mass so much as its radius, which is not the case in radial velocity surveys, in which the planet's radius does not play a role in determining the detectability of the object. \citet{fortney2007} shows that 1.0-1.2 $R_J$ objects can span three decades in mass, from nearly stellar to down to super-earth masses, depending on the composition of the planet. Radial velocity surveys, on the other hand, are sensitive to the mass of the object, and can span about an order of magnitude in radius for different compositions. For most ground based transit surveys and typical radial velocity precisions both methods tend to probe Jupiter-like objects, but it is important to note that the populations probed in each survey method can differ. For definiteness, we will assume that all planets detected in RV surveys with $M\sin{i}>0.1$ $M_J$ have radii $ \gtrsim R_J$

Additionally, we must account for the metallicity biases inherent to each survey method. To better compare our results to those of other surveys we must normalize our upper limits to the metallicities typical of the comparison surveys. FV05 performed a careful analysis of stellar metallicities in Keck, Lick, and Anglo-Australian Observatory radial velocity surveys, and they find that the mean metallicity of 1040 planet search stars is 0.09 dex more metal-rich than volume-limited sample selected from the same dataset. Furthermore, stars that bear planets are 0.226 dex more metal rich than the volume-limited sample. The average metallicity of planet-bearing stars in FV05 is $\textrm{[Fe/H]} \approx 0.142$ which implies a mean metallicity for the RV planet search stars of $\textrm{[Fe/H]}=0.142-0.226+0.09=0.006$ and $\textrm{[Fe/H]} = 0.142-0.226=-0.084$ for the volume-limited sample. We therefore adopt $\textrm{[Fe/H]} = 0.006$ as the average metallicity of the stars used in C08 to estimate the planet frequency in the Keck survey, which based on a target sample similar to that analyzed in FV05.

In the case of the OGLE III survey, there is only a very weak bias towards more metal rich stars relative to a volume limited survey (G06). We therefore adopt the FV05 average metallicity for the volume-limited sample as the mean metallicity of the OGLE stars. Although the FV05 volume-limited sample only extends out to 20 pc while the OGLE survey detects much more distant stars, we assume that the FV05 volume-limited average metallicity is a fair estimate of the average OGLE metallicity. The uncertainties in the determination of the cluster metallicities are far larger than those implicit in this assumption, and so for our purposes, $\textrm{[Fe/H]} = -0.084$ is a sufficiently accurate estimate of the average OGLE field star metallicities.

We cannot better normalize the \citet{weldrake2008} and \citet{paulson2004} results. \citet{paulson2004} does not calculate explicit upper limits or detection efficiencies, which makes comparison difficult. \citet{weldrake2008} focuses on the globular cluster $\omega$ Cen which has $\textrm{[Fe/H]} < -0.5$, for which the FV05 relation no longer holds. Therefore, although we quote the results of these surveys for completeness, we cannot directly compare them to our results.

\begin{figure}

\centerline{\includegraphics[scale = 0.4]{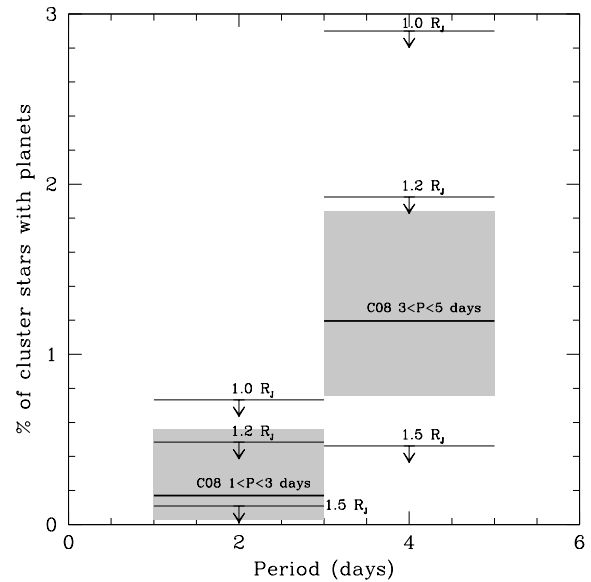}}
\caption{Upper limits derived using all six surveys. Upper limits are quoted at a metallicity [Fe/H] =+0.006, the mean metallicity of stars in RV planet search surveys according to FV05. The dark lines and shaded regions show the C08 short-period planet frequencies and 68\% confidence intervals, respectively.}
\label{fig:cumming_compare}
\end{figure}

\begin{figure}
\centerline{\includegraphics[scale = 0.4]{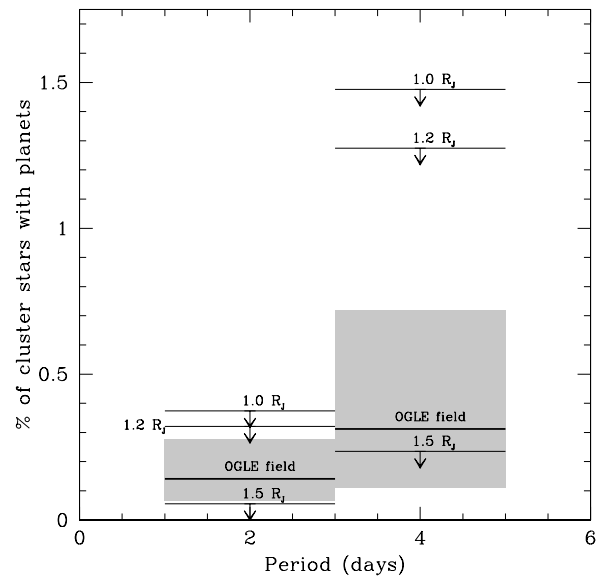}}
\caption{Upper limits from the 6 main surveys compared to the frequencies of VHJs and HJ found in \citet{gould2006}. Upper limits are quoted at $90\%$ confidence and  [Fe/H] = -0.084, which corresponds to the metallicity of the volume-limited sample in FV05. \citet{gould2006} frequencies and the corresponding 68\% confidence regions are shown as dark lines and shaded regions, respectively. }
\label{fig:OGLE_compare}
\end{figure}

We can quote our upper limits normalized to the mean metallicities of the C08 radial velocity surveys and the OGLE transit surveys. For $\langle\textrm{[Fe/H]}\rangle = 0.006$, corresponding to radial velocity surveys, we find upper limits of 2.9\%, 1.9\% and 0.46\% for 1.0, 1.2, and 1.5$R_J$ HJs, respectively. For VHJs the limits are 0.73\%, 0.48\%, and 0.11\% for 1.0, 1.2, and 1.5$R_J$ respectively. For an average metallicity or $\textrm{[Fe/H]} = -0.084$, corresponding to the OGLE III surveys, we find upper limits of the planet fraction of HJs at $90\%$ confidence (as in OGLE) 1.5\% of  1.3\% and 0.24\% for 1.0, 1.2, and 1.5 $R_J$ planets respectively. Similarly, for VHJs we find limits of 0.37\%, 0.32\%, and 0.06\% for 1.0, 1.2, and 1.5 $R_J$ planets respectively (The frequencies quoted in G06 were $\sim0.31\%$ and $\sim0.14\%$ for HJ and VHJ respectively for 1.0-1.25 $R_J$ planets). Figures \ref{fig:cumming_compare} and \ref{fig:OGLE_compare} graphically display these results.

For each comparison, our upper limits are consistent with the results of other surveys. Our derived upper limits for 1.0 and 1.2 $R_J$ lie above the short-period planet frequencies derived using both the OGLE transit surveys and the C08 result for the Keck Planet Search. Another way to see this is to ask: how many planets should we expect to have seen in these 6 open cluster surveys given the frequency of short-period giants published in G06 and C08, given the known detection efficiencies and number of cluster stars targeted by each survey? The answer is 0.86, 1.00, and 5.80 for 1.0, 1.2, and 1.5 $R_J$ VHJs, respectively, and 0.48, 0.56, and 3.03 for 1.0, 1.2, and 1.5 $R_J$ HJs using the G06 frequencies. Using the C08 frequencies we estimate the surveys should have detected 0.70, 1.06, and 4.69 planets for 1.0, 1.2, and 1.5 $R_J$ VHJs and 1.24, 1.86, and 7.75 planets for HJs. The probability of a non-detection when one expects to detect 1.86 1.2 $R_J$ planets is 15.5\%. Although it is possible that these surveys could yield a non-detection for a planet frequency equal to that found in RV surveys, the probablity is relatively low, suggesting that these transit survey results are approaching interesting constraints. However, as we discuss in \S \ref{sec:future_surveys}, the number of additional observed stars needed to significantly improve these existing upper limits becomes prohibitively large. 

The expected number of planets is only large in the case of 1.5 $R_J$ objects, and both the G08 and C06 frequencies are not directly comparable to planets in this radius range. For the OGLE surveys, this is because the upper limits quoted in \citet{gould2006} are for 1.0-1.25 $R_J$ planets; no planets were found with radii larger than this, and any limit on 1.5 $R_J$ planets from the OGLE survey would be tighter. In the case of the C08 result, one must recall that radial velocity surveys probe planetary mass, not radius. It is likely the case that the planets with $M\sin{i}>0.1$ $M_J$ detected in RV surveys are of smaller radii. The upper limits for a more likely average radius of 1.2 $R_J$ for such planets, are \emph{not} in conflict with the C08 frequencies.

\subsection{RV detected planets in open clusters}

We have evidence from other methods of planet detection that planets do exist in open clusters: \citet{lovis2007} reported the RV detection of a massive planet with a minimum mass of 10.6 $M_{J}$ on a 714 day orbit around a red giant in the open cluster NGC 2423. Likewise, \citet{sato2007} reported the discovery of a planet in the Hyades open cluster with $m\sin{i}=7.6\pm0.2$ $M_{J}$ and period $594.9\pm5.3$ days, again using radial velocities. Although these planets have much longer periods than any of those to which these transit surveys would be sensitive, they still represent evidence that open clusters are not devoid of massive planetary bodies.
\begin{deluxetable*}{llllll}
\tabletypesize{\footnotesize}
\tablecolumns{6}
\tablewidth{0pt}
\tablecaption{Additional Photometric Surveys}
\tablehead{
                       \colhead{Cluster}&
                       \colhead{[Fe/H]\tablenotemark{a}}&
                       \colhead{stars with}&
                       \colhead{cluster stars}&
                       \colhead{method of membership}&
                       \colhead{Reference}\\
                       \colhead{}&
                       \colhead{}&
                       \colhead{$\sim1\%$ phot.}&
                       \colhead{}&
                       \colhead{estimation}&
                       \colhead{}}

\startdata
NGC 2301 & +0.06 & 4000   & $\sim1500$ & estimate from CMD\tablenotemark{b}                   & \citet{howell2005}\\
NGC 2632 & +0.14 & \nodata& $\sim150$  & WEBDA memberships                                    & \citet{pepper2008}\\
NGC 2660 & -0.18 & 3000   & $\sim600$  & comparison fields                                    & \citet{vonbraun2004}\\
        &       &        &            &                                                      & \citet{vonbraun2005}\\
NGC 6208 &\nodata& 5000   & $\sim1000$ & comparison fields                                    & \citet{lee2004}\\
NGC 6633 &\nodata& 2000   & $\sim600 $ & by-eye estimate\tablenotemark{b}                     & \citet{hidas2005}\\
NGC 6819 & +0.07 & 11500  & $\sim700$  & colors                                               & \citet{street2003}\\
NGC 6940 & +0.01 & 4400   & $\sim100$  & most cluster members saturated                       & \citet{hood2005}\\
NGC 7086 &\nodata& 445    & $\sim150$  & Besanc\c{o}n model                                   & \citet{rosvick2006}\\
NGC 7789 & -0.08 & 2400   & $\sim240$  & star counts on CCD chips\tablenotemark{b}            & \citet{bramich2005}\\
NGC 6253\tablenotemark{c} & +0.36 & ?      &      ?     &                                                     & \citet{montalto2009}\\
\enddata
\tablenotetext{a} {Metallicities courtesy of WEBDA: http://www.univie.ac.at/webda/}
\tablenotetext{b} {Even rough estimates of the membership were not mentioned in the the reference, and we estimated membership ourselves. In all other cases the references had at least some, however rough, estimate for the number of cluster stars observed.}
\tablenotetext{c} {\citet{montalto2009} contained no information about the quality of the light curves or details about any planetary transit searches.}
\label{tbl:other_surveys}
\end{deluxetable*}

\subsection{What about all of the other surveys?}

We have chosen a subset of 6 transit surveys from $\sim20$ such projects because these surveys provided us with sufficient information to combine their results quantitatively, as described in \S\ref{sec:surveys}. However, this means that we have neglected the majority of the photometric surveys of open clusters in the literature. Here we attempt to ascertain how the addition of these surveys might alter our conclusions.

Although the remaining surveys do not meet our criteria for selection, we do our best to estimate to what degree these additional data affect the upper limits on the planet frequency.  In most of these surveys the authors made no attempt to put upper limits on the frequency of planets in the fields they observed, often selected transits by eye, and rarely estimated the number of cluster members observed.  We estimate from the contents of each paper the number of stars with sufficiently precise photometry for planetary transits to be visible, and the number of cluster members likely to be within this subset of stars. A list of the additional surveys and estimates of the number of stars they contribute to the total sample of cluster stars is presented in Table \ref{tbl:other_surveys}. In total, the 10 additional surveys add $\sim5000$ stars to our sample. Assuming an average detection efficiency of 0.5\%, and a maximum efficiency of 1.0\%, this implies that $f_{p_{other}} \approx \frac{3.0}{(5000)(0.01)} \le 6\%$ in the best case and $f_{p_{other}} \approx \frac{3.0}{(5000)(0.005)} \le 12\%$ on average. The constraints on the population of 1.0 $R_J$ HJ planets from the main 6 surveys are the weakest of any considered here with $f_p \le 0.055$. Using our standard method of combining upper limits, the addition of the 5000 stars from the other surveys yields a new limit of $f_p \le 0.029$, which is a $\sim50\%$ tighter constraint. In the case in which $\mathcal{P}_{\epsilon} = 0.005$, the combined upper limit is $f_p \le 0.037$. In the VHJ range, the upper limits on $f_p$ for 1.0 $R_J$ planets decreases by $18\%$ and $10\%$ for detection efficiencies of $1.0\%$ and $0.5\%$ respectively. At larger planetary radii the changes in $f_p$ for 1.5 $R_J$ planets due to the addition of these 5000 stars is of order $\sim5-20\%$. The conclusions do not qualitatively change with the inclusion of metallicity information. The mean weighted metallicity of the stars in the unused 10 surveys is $\langle\textrm{[Fe/H]}\rangle\approx +0.1$ (according to the
WEBDA database), comparable to  $\langle\textrm{[Fe/H]}\rangle\approx +0.09$ found for the 6 surveys utilized in this paper.

We expect that by neglecting the observations present in these other surveys our derived upper limits may be up to a factor of two too high. However, it is highly unlikely that all of these surveys have achieved the optimistic $1.0\%$ detection efficiency, and improbable that all have even achieved a $0.5\%$ efficiency. We therefore conclude that neglecting these other surveys does not qualitatively change our primarily conclusion, that the lack of detections is consistent with the hypothesis that cluster and field stars host the same planet population. We do note that, had each of these surveys carefully quantified their detection efficiencies, the constraints on the number of short-period planets in open clusters could have been noticeably tighter.

\subsection{Future transit surveys in open clusters: how many more stars are needed?}\label{sec:future_surveys}
It is useful to ask, given our results, how many more cluster stars must be observed in surveys with null results before the combined upper limit derived using all extant surveys becomes inconsistent with the observed frequency of planets around field stars. Because of the low detection efficiencies and relatively few suitable open clusters in our galaxy, it is unlikely that upper limits derived from transit surveys in open clusters will be inconsistent with the results of C08 and G06 in the near future, even if it is the case that the frequency of planets in open clusters is significantly lower than that of the field. For the C08 result, a combined upper limit (at 95\% confidence) would be inconsistent with the 68\% confidence lower bound on the C08 frequency of the fraction of stars with planets was found to be $f_p \le 0.0076$ for 1.2 $R_J$ planets in the $3<P<5$ day period range. The six-survey upper limit is $f_p \le 0.019$, which implies that $\sim 48,000$ more cluster stars must be observed (with null results) at a 0.5\% detection efficiency before the limit is in conflict with C08 (currently, $\sim13000$ stars have been observed at $\sim0.5\%$ efficiency). This number drops to 9000 with a detection efficiency of 1.0\%. Given that there are about 1200 known open clusters in the galaxy (WEBDA), only a handful of which are rich enough to contain thousands of stars, the only way to make the upper limits on the planet frequency in open clusters derived from transit surveys competitive with field surveys is to drastically increase the detection efficiency of the surveys \emph{and} number of clusters observed. When we ask how many stars would need to be observed to be competitive with the G06 field star planet frequency, the requirements are even more strict. Surveys would need to cover $\sim 387,000$ stars at 0.5\% detection efficiency before the upper limits, at 90\% confidence, would be inconsistent with the 68\% confidence lower bound on the G06 frequency. When we consider the upper limits on VHJs even more additional stars are required.

\section{Conclusion}\label{sec:conclusion}

We conclude that the current upper limits of the HJ and VHJ frequencies determined using the null results of transit surveys in open clusters do not suggest a significant difference between the frequency of planets in open clusters and the field. Open clusters remain, to the best of our knowledge, a viable and useful target for exoplanet transit surveys, and we do not yet have enough data to discern whether the environments of open clusters have any noticeable effect on planet formation and survival.  We recommend that any future surveys carefully quantify any null results, since the combination of many such outcomes has the potential to better constrain the frequency of short-period planets and answer interesting questions about the formation of planets in stellar clusters.

\section{Acknowledgments}
 We would like to thank K.Z. Stanek for the initial idea to do this analysis, and useful discussions throughout the process. We also thank Kenneth Janes, Thomas Beatty, Benjamin Shappee and Calen Henderson for useful discussions and comments on the manuscript. This research has made use of the WEBDA database, operated at the Institute for Astronomy of the University of Vienna, and the Exoplanet Orbit Database and the Exoplanet Data Explorer at exoplanets.org.

\end{document}